\documentclass[12pt]{article} 
\usepackage{graphicx}
\begin{document}
\baselineskip 10mm

\centerline{\bf Entanglement and quantum state engineering }
\centerline{\bf in the optically driven two-electron double-dot
structure}

\vskip 2mm

\centerline{Alexander V. Tsukanov}

\vskip 2mm

\centerline{\it Institute of Physics and Technology, Russian
Academy of Sciences}
\centerline{\it Nakhimovsky pr. 34, Moscow
117218, Russia}
\centerline{\it E-mail: tsukanov@ftian.oivta.ru}

\vskip 4mm

\begin{quotation}

We study theoretically the quantum dynamics of two interacting
electrons in the symmetric double-dot structure under the
influence of the bichromatic resonant pulse. The state vector
evolution is studied for two different pulse designs. It is shown
that the laser pulse can generate the effective exchange coupling
between the electron spins localized in different dots. Possible
applications of this effect to the quantum information processing
(entanglement generation, quantum state engineering) are
discussed.

\end{quotation}

\vskip 4mm

PACS: 03.67.Mn, 71.70.Gm, 73.21.La, 73.23.-b

\vskip 6mm

\newpage

\centerline{\bf I. INTRODUCTION}

 In recent years, the low-dimensional semiconductor structures
containing a small number of electrons in the size-quantized
conduction band have attracted much attention. The main reason for
that interest is the progress in up-to-date technology that allows
one to fabricate the nanostructures with high precision \cite{1}.
Along with the use of advanced techniques of population control
\cite{2}, it makes the creation of macroatoms with desired
properties a standard experimental tool at hand. There is a lot of
possible applications of this field of solid state physics
\cite{3}. One of them is concerned with rapidly developing quantum
computing and quantum information processing \cite{4}.

 The key point of implementation of quantum computer's
hardware is to find an appropriate physical system characterized
by the well-defined Hilbert space that would allow for efficient
external (classical) control. Among the systems proposed for
operation with the quantum information, a two-level system (qubit)
is the most studied one. For practical applications, one has to
look for a physical system that could serve as a base for a {\it
scalable} quantum computer. Only such scalable quantum computers
would outperform their classical counterparts
 for several classes of computational problems. It is commonly believed that the problem of
scalability can be effectively solved with solid-state systems.
Many proposals for solid-state qubit realization have been made.
Here we mention the superconducting devices using a Cooper-pair
box \cite{5}, phosphorous donors embedded in a silicon host
\cite{6,7,8}, and a wide class of the systems based on the quantum
dots (QDs) (see, e. g., \cite{9,11,12,13,17,19,20,21,22,23}).

 Loss and DiVincenzo have proposed to encode the quantum
information into electron spins contained in laterally coupled
electrostatically-formed QDs \cite{9}. In their model, the local
alternating magnetic field is used for single-qubit rotations,
while the electrostatic gate voltage brings about the interqubit
coupling. Besides, the static magnetic field can be applied to the
double-dot structure to adjust the exchange coupling between
electrons. Upon the control of the exchange
 during the voltage pulse action, the logic operations on the
neighboring electron spins become possible. To organize a
non-trivial two-qubit gate (e. g., XOR) one needs to perform the
"square-root of swap" operation combined with appropriate
single-qubit rotations \cite{10}. Furthermore, the exchange
control allows one to transform the product state of the qubits
into the non-separable superposition of the qubit states which is
the highly-entangled state \cite{47,48,49,50}.

 Here we present a novel scheme for manipulation with the quantum
states of two interacting electrons in the double-dot structure.
It is based on the interaction of electrons with the coherent
electromagnetic pulse. We are interested in the design of the
external driving field that allows to achieve the reliable control
on the quantum dynamics of the system. As we shall show, this may
be done by the generalization of the driving schemes developed for
a single electron confined in the double-dot structure
\cite{13,14,15,16,18}. The sequence of laser pulses, instead of
electrostatic ones, is applied to the double-dot structure, and
the all-optical quantum state engineering is realized via the
optically induced transitions between the size-quantized
two-electron levels. The resonant character of the electron-pulse
interaction provides the high selectivity of the corresponding
transitions, thus making the scheme robust against the unwanted
excitations far from the resonance. We consider the resonant
transitions between two degenerate ground states and the auxiliary
excited states. We use the ground (localized) states of the system
as qubit states. This allows us to isolate them from each other
after the pulse is off. Besides, the scheme permits one to operate
with the strongly-detuned pulses in the Raman-like regime, where
one (or several) auxiliary state remains unpopulated. This seems
to be important for the structures with the strong decoherence. We
shall see that two resonant pulses are sufficient for
implementation of such a rotation in the subspace spanned by the
ground states that corresponds to generation of the entanglement
of electron spins. Thus, the two-electron double-dot structure may
be exploited as the spin entangler.

 The paper is organized as follows. In Sec. II we study the two-electron double-dot
structure interacting with the external pulse. The energy spectrum
and the stationary eigenstates of the structure are obtained from
the analysis of the extended Hubbard model. The probability
amplitudes of the eigenstates relevant for the quantum dynamics
are found from the solution of the non-stationary Schr\"odinger
equation. In Sec. III we describe the quantum operations that may
be realized in this system under the influence of the laser pulses
and propose the scheme for the entanglement generation. The
results are summarized in Sec. IV.

\vskip 5mm

\centerline{\bf II. THE MODEL}

\centerline{\bf A. The eigenstates and the eigenenergies of the
stationary Hamiltonian}

 We consider the double-dot structure (see Fig.1)
 containing two interacting electrons in the size-quantized conduction band. For
the sake of simplicity, we suppose the dots $A$ and $B$ to be
identical. The existence of at least two one-electron orbital
states $\left| {A\left( B \right)0} \right\rangle $ and $\left|
{A\left( B \right)1} \right\rangle $ (ground and excited) in each
of the QDs is assumed, with the one-electron wave functions
$\varphi _{A(B)0} \left( {\bf{r}} \right)=\left\langle {{\bf{r}}}
 \mathrel{\left | {\vphantom {{\bf{r}} {A(B)0 }}}
 \right. \kern-\nulldelimiterspace}
 {{A(B)0 }} \right\rangle
$ and $\varphi _{A(B)1} \left( {\bf{r}}\right )=\left\langle
{{\bf{r}}}
 \mathrel{\left | {\vphantom {{\bf{r}} {A(B)1 }}}
 \right. \kern-\nulldelimiterspace}
 {{A(B)1 }} \right\rangle $, respectively.
Provided that the distance between the QDs is sufficiently large,
the wave functions of the QD ground states are localized in
corresponding QDs, and their overlap can be neglected. The overlap
between the ground state and the excited state belonging to
different QDs will be neglected as well: $\left\langle {{A(B)0 }}
 \mathrel{\left | {\vphantom {{A(B)0 } {B(A)1 }}}
 \right. \kern-\nulldelimiterspace}
 {{B(A)1 }} \right\rangle \approx 0$. The excited levels are
chosen to be close to the edge of the potential barrier separating
the QDs. They couple through the electron tunneling \cite{13}.

 Following the procedure used in Ref. \cite{25}, we transform
 the one-electron orbitals of the isolated QDs into the orthonormal
one-electron orbitals, accounting for the hybridization of the
excited states: $\tilde \varphi _{A0} \left( {\bf{r}} \right)$ = $
\varphi _{A0} \left( {\bf{r}} \right)$, $\tilde \varphi _{B0}
\left( {\bf{r}} \right)$ = $\varphi _{B0} \left( {\bf{r}}
\right)$, $\tilde \varphi _{A1} \left( {\bf{r}} \right)$ = $
\left( {{1 \mathord{\left/
 {\vphantom {1 {2\sqrt {1 + s} \,\,\, + }}} \right.
 \kern-\nulldelimiterspace} {2\sqrt {1 + s} \,\,\, + }}\,\,{1 \mathord{\left/
 {\vphantom {1 {2\sqrt {1 - s} \,}}} \right.
 \kern-\nulldelimiterspace} {2\sqrt {1 - s} \,}}} \right)\varphi _{A1}$$ \left( {\bf{r}} \right) + \left( {{1 \mathord{\left/
 {\vphantom {1 {2\sqrt {1 + s} \,\,\, - }}} \right.
 \kern-\nulldelimiterspace} {2\sqrt {1 + s} \,\,\, - }}\,\,{1 \mathord{\left/
 {\vphantom {1 {2\sqrt {1 - s} \,}}} \right.
 \kern-\nulldelimiterspace} {2\sqrt {1 - s} \,}}} \right)$$\varphi _{B1} \left( {\bf{r}} \right)$,
  $\tilde \varphi _{B1} \left( {\bf{r}} \right)$ =
$ \left( {{1 \mathord{\left/
 {\vphantom {1 {2\sqrt {1 + s} \,\,\, + }}} \right.
 \kern-\nulldelimiterspace} {2\sqrt {1 + s} \,\,\, + }}\,\,{1 \mathord{\left/
 {\vphantom {1 {2\sqrt {1 - s} \,}}} \right.
 \kern-\nulldelimiterspace} {2\sqrt {1 - s} \,}}} \right)\varphi _{B1} \left( {\bf{r}} \right) + \left( {{1 \mathord{\left/
 {\vphantom {1 {2\sqrt {1 + s} \,\,\, - }}} \right.
 \kern-\nulldelimiterspace} {2\sqrt {1 + s} \,\,\, - }}\,\,{1 \mathord{\left/
 {\vphantom {1 {2\sqrt {1 - s} \,}}} \right.
 \kern-\nulldelimiterspace} {2\sqrt {1 - s} \,}}} \right)$ $\varphi _{A1} \left( {\bf{r}} \right)$
. Here $s =\left\langle {{A1 }}
 \mathrel{\left | {\vphantom {{A1 } {B1 }}}
 \right. \kern-\nulldelimiterspace}
 {{B1 }} \right\rangle $
  is the overlap between the excited states of the QDs.

 The Hamiltonian of two interacting electrons confined in the symmetric double-dot
structure is $H_0 = h\left( {{\bf{r}}_1 } \right) + h\left(
{{\bf{r}}_2 } \right) + w\left( {\left| {{\bf{r}}_1  - {\bf{r}}_2
} \right|} \right)$ (\cite{10,25}),  where $h\left( {{\bf{r}}_i }
\right)$ is the one-particle Hamiltonian that includes the kinetic
and potential terms,
 and $w\left( {\left| {{\bf{r}}_1  - {\bf{r}}_2 } \right|} \right)$ is the Coulomb interaction between
the electrons.
   We consider the case that the electrons have the opposite spins. The
Hamiltonian can be expressed in terms of the extended Hubbard
model \cite{25}:

 \begin{equation}
 \begin{array}{l}
 H_{0} = \sum\limits_{\rm{\sigma }} {\left[ {\varepsilon _{\rm{0}} \left( {n_{A0,{\rm{\sigma }}}  + n_{B0,{\rm{\sigma }}} } \right) + \varepsilon _{\rm{1}} \left( {n_{A1,{\rm{\sigma }}}  + n_{B1,{\rm{\sigma }}} } \right)} \right]}  - \tilde t\sum\limits_{\rm{\sigma }} {\left( {a_{A1,{\rm{\sigma }}}^{\rm{ + }} a_{B1,{\rm{\sigma }}}  + h.c.} \right)}  +  \\
 \,\,\,\,\,\, + \sum\limits_{{\rm{\sigma }}\ne{\rm{\sigma '}}} {\left[ {V_{00} n_{A0,{\rm{\sigma }}} n_{B0,{\rm{\sigma '}}}  + V_{01} \left( {n_{A0,{\rm{\sigma }}} n_{B1,{\rm{\sigma '}}}  + n_{B0,{\rm{\sigma }}} n_{A1,{\rm{\sigma '}}} } \right) + V_{11} n_{A1,{\rm{\sigma }}} n_{B1,{\rm{\sigma '}}} } \right]}  +  \\
 \,\,\,\,\,\, + U_{00} \left( {n_{A0, \uparrow } n_{A0, \downarrow }  + n_{B0, \uparrow } n_{B0, \downarrow } } \right) + U_{11} \left( {n_{A1, \uparrow } n_{A1, \downarrow }  + n_{B1, \uparrow } n_{B1, \downarrow } } \right) +  \\
 \,\,\,\,\,\, + U_{01} \sum\limits_{{\rm{\sigma }}\ne{\rm{\sigma '}}} {\left( {n_{A0,{\rm{\sigma }}} n_{A1,{\rm{\sigma '}}}  + n_{B0,{\rm{\sigma }}} n_{B1,{\rm{\sigma '}}} }
 \right),}\\
 \end{array}
\end{equation}
where $a_{k,{\rm{\sigma }}}^{\rm{ + }} $ creates an electron in
the state with the wave function ${\tilde \varphi _k \left(
{\bf{r}} \right)}$
 ($k=A0,\ B0,\ A1$, and $\ B1$) and the spin $\sigma=+1/2,
-1/2$; $n_{k,{\rm{\sigma }}} = a_{k,{\rm{\sigma }}}^{\rm{ + }}
a_{k,{\rm{\sigma }}}^{} $ is the particle number operator acting
on the state vectors in the occupation number representation;
$\varepsilon _0  = \int {\tilde \varphi _{A\left( B \right)0}^*
\left( {\bf{r}} \right)h\left( {\bf{r}} \right)\tilde \varphi
_{A\left( B \right)0} \left( {\bf{r}} \right)} d{\bf{r}}$ and
$\varepsilon _1  = \int {\tilde \varphi _{A\left( B \right)1}^*
\left( {\bf{r}} \right)h\left( {\bf{r}} \right)\tilde \varphi
_{A\left( B \right)1} \left( {\bf{r}} \right)} d{\bf{r}}$ are the
one-electron energies of the ground and excited states,
respectively (the same for both QDs); $\tilde t = \int {\tilde
\varphi _{A1}^* \left( {\bf{r}} \right)h\left( {\bf{r}}
\right)\tilde \varphi _{B1} \left( {\bf{r}} \right)} d{\bf{r}}$
 is
the matrix element for the electron hopping between the excited
states of the QDs; $V_{00}  = V_{A0,\,B0} $, $V_{01}  =
V_{A0,\,B1} = V_{A1,\,B0} $, $V_{11}  = V_{A1,\,B1} $, $U_{01}  =
U_{A0,\,A1} = U_{B0,\,B1} $ are the Coulomb interaction energies
for electrons occupying different states, and $U_{00}  =
U_{A0,\,A0} = U_{B0,\,B0} $, $U_{11} = U_{A1,\,A1} = U_{B1,\,B1} $
are the Coulomb interaction energies for electrons occupying the
same orbital state. For the sake of simplicity we have omitted in
Eq. (1) the higher-order terms such as the cotunneling, the direct
exchange, e.t.c. In what follows, we consider the system in the
strong-confinement regime where the Coulomb correlations are small
compared to the level spacing, i. e., $V_{00} ,\,V_{01} ,\,V_{11}
\, <  < U_{00} ,\,U_{01} ,\,U_{11} < < \varepsilon _1 -
\varepsilon _0 \,$.

 The straightforward diagonalization of the Hamiltonian results in
the two-electron eigenstates $\left| n \right\rangle $ $\left( {n
= 1 - 16} \right)$ that can be expressed in terms of the four-site
basis vectors $\left| {n_{A0,\sigma } } \right\rangle \otimes
\left| {n_{B0,\sigma '} } \right\rangle \otimes \left|
{n_{A1,\sigma ''} } \right\rangle  \otimes \left| {n_{B1,\sigma
'''} } \right\rangle = \left| {n_{A0,\sigma } ,n_{B0,\sigma '}
,n_{A1,\sigma ''} ,n_{B1,\sigma '''} } \right\rangle $ as follows:

\begin{equation}
\begin{array}{l}
 \left| 1 \right\rangle  = \left| {1_ \uparrow  ,1_ \downarrow  ,0,0} \right\rangle ,\,\,\left| 2 \right\rangle  = \left| {1_ \downarrow  ,1_ \uparrow  ,0,0} \right\rangle ,\,\,\left| 3 \right\rangle  = \left| {2,0,0,0} \right\rangle ,\,\,\left| 4 \right\rangle  = \left| {0,2,0,0} \right\rangle , \\
 \left| 5 \right\rangle  = C_ +  \left[ {\left( {v + \sqrt {v^2  + \tilde t^2 } } \right)\left| {1_ \uparrow  ,0,0,1_ \downarrow  } \right\rangle  + \tilde t\left| {1_ \uparrow  ,0,1_ \downarrow  ,0} \right\rangle } \right], \\
 \left| 6 \right\rangle  = C_ +  \left[ {\left( {v + \sqrt {v^2  + \tilde t^2 } } \right)\left| {0,1_ \downarrow  ,1_ \uparrow  ,0} \right\rangle  + \tilde t\left| {0,1_ \downarrow  ,0,1_ \uparrow  } \right\rangle } \right], \\
 \left| 7 \right\rangle  = C_ +  \left[ {\left( {v + \sqrt {v^2  + \tilde t^2 } } \right)\left| {0,1_ \uparrow  ,1_ \downarrow  ,0} \right\rangle  + \tilde t\left| {0,1_ \uparrow  ,0,1_ \downarrow  } \right\rangle } \right], \\
 \left| 8 \right\rangle  = C_ +  \left[ {\left( {v + \sqrt {v^2  + \tilde t^2 } } \right)\left| {1_ \downarrow  ,0,0,1_ \uparrow  } \right\rangle  + \tilde t\left| {1_ \downarrow  ,0,1_ \uparrow  ,0} \right\rangle } \right], \\
 \left| 9 \right\rangle  = C_ -  \left[ {\left( {v - \sqrt {v^2  + \tilde t^2 } } \right)\left| {1_ \uparrow  ,0,0,1_ \downarrow  } \right\rangle  + \tilde t\left| {1_ \uparrow  ,0,1_ \downarrow  ,0} \right\rangle } \right], \\

 \left| {10} \right\rangle  = C_ -  \left[ {\left( {v - \sqrt {v^2  + \tilde t^2 } } \right)\left| {0,1_ \downarrow  ,1_ \uparrow  ,0} \right\rangle  + \tilde t\left| {0,1_ \downarrow  ,0,1_ \uparrow  } \right\rangle } \right], \\
 \left| {11} \right\rangle  = C_ -  \left[ {\left( {v - \sqrt {v^2  + \tilde t^2 } } \right)\left| {0,1_ \uparrow  ,1_ \downarrow  ,0} \right\rangle  + \tilde t\left| {0,1_ \uparrow  ,0,1_ \downarrow  } \right\rangle } \right], \\
 \left| {12} \right\rangle  = C_ -  \left[ {\left( {v - \sqrt {v^2  + \tilde t^2 } } \right)\left| {1_ \downarrow  ,0,0,1_ \uparrow  } \right\rangle  + \tilde t\left| {1_ \downarrow  ,0,1_ \uparrow  ,0} \right\rangle } \right], \\
 \left| {13} \right\rangle  = \tilde C_ +  \left[ {\left( {\tilde v + \sqrt {\tilde v^2  + 4\tilde t^2 } } \right)\frac{{\left| {0,0,1_ \uparrow  ,1_ \downarrow  } \right\rangle  + \left| {0,0,1_ \downarrow  ,1_ \uparrow  } \right\rangle }}{{\sqrt 2 }} + 2\tilde t\frac{{\left| {0,0,2,0} \right\rangle  + \left| {0,0,0,2} \right\rangle }}{{\sqrt 2 }}} \right], \\
 \left| {14} \right\rangle  = \frac{{\left| {0,0,1_ \uparrow  ,1_ \downarrow  } \right\rangle  - \left| {0,0,1_ \downarrow  ,1_ \uparrow  } \right\rangle }}{{\sqrt 2 }},\,\,\left| {15} \right\rangle  = \frac{{\left| {0,0,2,0} \right\rangle  - \left| {0,0,0,2} \right\rangle }}{{\sqrt 2 }}, \\
 \left| {16} \right\rangle  = \tilde C_ -  \left[ {\left( {\tilde v - \sqrt {\tilde v^2  + 4\tilde t^2 } } \right)\frac{{\left| {0,0,1_ \uparrow  ,1_ \downarrow  } \right\rangle  + \left| {0,0,1_ \downarrow  ,1_ \uparrow  } \right\rangle }}{{\sqrt 2 }} + 2\tilde t\frac{{\left| {0,0,2,0} \right\rangle  + \left| {0,0,0,2} \right\rangle }}{{\sqrt 2 }}} \right], \\
 \end{array}
\end{equation}

where

\begin{equation}
C_ \pm   = \frac{1}{{\sqrt {\tilde t^2  + \left( {v \pm \sqrt {v^2
+ \tilde t^2 } } \right)^2 } }}, \,\,\,\,\,\,\,\,\\\\\ \tilde C_
\pm = \frac{1}{{\sqrt {4\tilde t^2  + \left( {\tilde v \pm \sqrt
{\tilde v^2  + 4\tilde t^2 } } \right)^2 } }},
\end{equation}

and

\begin{equation}
u = \frac{{U_{01}  + V_{01} }}{2},\,\,\,v = \frac{{U_{01} - V_{01}
}}{2},\,\,\,\tilde u = \frac{{U_{11}  + V_{11} }}{2},\,\,\,\tilde
v = \frac{{U_{11}  - V_{11} }}{2}.
\end{equation}

The corresponding eigenenergies $E_n$ ($n=1-16$) are $E_{1} =
E_{2} =2\varepsilon _{0} + V_{00}$, $E_{3}=E_{4}=2\varepsilon _0 +
U_{00}$, $E_{5}=E_{6}=E_{7}=E_{8}= \varepsilon _0 + \varepsilon _1
+ u - \sqrt {v^2  + \tilde t^2 }$, $E_{9}=E_{10}=E_{11}=E_{12}=
\varepsilon _0 + \varepsilon _1 + u + \sqrt {v^2  + \tilde t^2 }$,
$E_{13}=2\varepsilon _1 + \tilde u - \sqrt {\tilde v^2  + 4\tilde
t^2 }$, $E_{14}=2\varepsilon _{1}  + V_{11}$, $E_{15}=2\varepsilon
_{1}  + U_{11}$, and $E_{16}=2\varepsilon _1 + \tilde u + \sqrt
{\tilde v^2  + 4\tilde t^2 }$.

 Since we suppose $V_{00} ,\,V_{01} ,\,V_{11}
\, <  < U_{00} ,\,U_{01} ,\,U_{11}  < < \varepsilon _1  -
\varepsilon _0,$ the coupling between the electrons gives rise to
three energy manifolds around the values of the energy of two
non-interacting electrons, $2\varepsilon _0 $, $\varepsilon _0 +
\varepsilon _1 $, and $2\varepsilon _1 $ .
 The eigenstates $\left| 1 \right\rangle$ and $ \left| 2
\right\rangle $ with the energy $\varepsilon _{00} =E_1=E_2$ are
the ground states of the two-electron system, where both electrons
are localized in the lowest orbital levels of different QDs. The
states $\left| 3 \right\rangle$ and $\left| 4 \right\rangle$,
where the electrons occupy the lowest orbital level in the same
QD, have the energy $\tilde \varepsilon _{00} = E_3=E_4$. Next
eight eigenstates correspond to the situation where one electron
is localized in one of the QDs and another electron is excited and
delocalized over the double-dot structure due to tunneling. Four
of those states, namely $\left| 5 \right\rangle,\ \left| 6
\right\rangle,\ \left| 7 \right\rangle$, and $\left| 8
\right\rangle$, constitute the subspace with the energy
$\varepsilon _{01}^ - = E_5=E_6=E_7=E_8$, whereas other four
states, i.e. $\left| 9 \right\rangle, \left| 10 \right\rangle,
\left| 11 \right\rangle$, and $\left| 12 \right\rangle$, form the
subspace with the energy $\varepsilon _{01}^ + =
E_9=E_{10}=E_{11}=E_{12}$. The states $\left| 13 \right\rangle,
\left| 14 \right\rangle, \left| 15 \right\rangle$, and $\left| 16
\right\rangle$ are the states where both electrons are excited.
Their properties have been studied in Refs. \cite{10,24,25,27,39}.
The eigenstate $\left| 13 \right\rangle$ is the singlet state with
the energy $\varepsilon _{11}^{ - s}  = E_{13}$. The triplet state
$\left| 14 \right\rangle$ has the energy $\varepsilon _{11}^{ t} =
E_{14}$. Two remaining eigenstates, $\left| 15 \right\rangle$ and
$\left| 16 \right\rangle$, are the singlet states with the
energies $\varepsilon _{11}^{ a}  = E_{15}$ and $\varepsilon
_{11}^{ + s} = E_{16}$, respectively. The superscript $s(a)$
denotes the symmetry (antisymmetry) of the corresponding state
under the spatial reflection operation.

 We see that the states $\left| 1 \right\rangle $ and $\left| 2
\right\rangle $ are degenerate. It follows from the fact that both
tunneling and exchange processes are prohibited due to the strong
confinement. Thus if the system was initialized, say, in the state
$\left| 1 \right\rangle $, we assert that an electron in the state
$\left| A0(B0) \right\rangle $ has the spin $\sigma =1/2 (-1/2)$
\cite{24}. Instead, the states $\left| 13 \right\rangle $ -
$\left| 16 \right\rangle $
 are hybridized, and electron spins in those states are correlated.
 In the next section, we shall work with the state $\left| {13} \right\rangle $.
  We shall show that together with the
doubly-occupied  states $\left| {3} \right\rangle $ and $\left|
{4} \right\rangle $, it can be used to entangle the electrons,
initially decoupled from each other. To achieve this goal, one
needs an effective coupling mechanism for the controlled
interaction between the electrons.

\centerline{\bf B. Laser-induced electron dynamics}

 The standard method usually considered for electron charge and
spin manipulations in a double-dot structure is based on the
electrostatic and magnetic field control. As was proposed in Ref.
\cite {10}, by applying an adiabatically switched voltage pulse
one can rise or lower the potential barrier between QDs. This
allows one to organize the controlled electron-electron
interaction and, as a consequence, the quantum state engineering.
The main difficulty inherent to this method lies in the trade-off
between the requirement of the small operation times, as compared
to the decoherence time, and the adiabatic character of the
electron tunneling process. In particular, the violation of
adiabatic conditions will lead to the double-occupancy of one of
the QDs, resulting in the quantum information leakage from the
computational subspace. To solve this problem, several approaches
have been developed, see Refs. \cite{26,27}.

  Here we propose an alternative scheme of the controlled quantum dynamics
  of the two-electron system. It is based on the resonant laser pulses that induce the transitions between
 the electron states in the double-dot structure. In recent works it was shown that
 similar techniques may be utilized to drive an excess electron
 between two QDs \cite{13,14,15,16}. For the exciton-based quantum computers the
 interqubit coupling schemes based on the coherent dynamics of localized \cite{23,28,29,30,31} and
 delocalized \cite{32,33} excitons under the influence of appropriately tuned laser pulses
 were developed. The superexchange coupling between two deep donor impurities in silicon
 mediated by the optically-excited electron of the control atom
 was discussed in Ref. \cite{34}. In our model we consider the electron dynamics involving
 the two-electron states (Eq. (2)) and show how to choose the pulse parameters
 to obtain the final quantum state with the desired properties.

Let our system be initially in the state $\left| 1 \right\rangle
$. To drive it into the superposition of the states $\left| 1
\right\rangle $ and $\left| 2 \right\rangle $, we make
 use of two resonant laser pulses that act simultaneously and
induce the optical dipole transitions between each of those states
and the auxiliary excited states. As we shall show, the proper
choice of the pulse parameters, such as the intensities, the
detunings of pulse frequencies from the resonance, and the pulse
durations, allows one to realize the inversion operation and
create the maximally entangled states, e. g., the Bell states.
There are several ways to achieve this goal. We study two of them,
 the most transparent in our opinion.

We consider the quantum evolution of the system under the
influence of two laser pulses. The model Hamiltonian has the form
\begin{equation}
 H = H_0  + \left[ V_1 \cos \left( {\omega _1 t + \varphi _1 } \right) + V_2 \cos \left( {\omega _2 t + \varphi _2 } \right)
 \right]
 \left[ {{\rm{\theta }}\left( t \right) - {\rm{\theta }}\left( {t
- T} \right)} \right] ,
\end{equation}
where $V_k  = - e{\bf{E}}_k \left( {{\bf{r}}_1 + {\bf{r}}_{\bf{2}}
} \right)$, $k=1, 2$; ${\bf{E}}_{k}$, ${\omega _k }$, and
${\varphi _k }$ are the amplitude, the frequency and the phase of
the $k$-th pulse, respectively, $e$ is the electron charge,
${{\rm{\theta }}\left( t \right)}$ is the step function
 and $T$ is the pulse duration.

 The state vector of the system may be represented in  terms of the
eigenstates $\left| 1 \right\rangle $ - $\left| 16 \right\rangle $
, Eqs. (2), of the stationary Hamiltonian $H_0$ as
\begin{equation}
 \left| {\Psi (t)} \right\rangle  = \sum\limits_{n = 1}^{16} {c_n(t) \exp \left( { -
i E _n t} \right)} \left| n
 \right\rangle
\end{equation}
(hereafter $\hbar  = 1$).

 The quantum evolution of the state vector under the influence of two laser pulses
is governed by the non-stationary Schr\"odinger equation
\begin{equation}
 i\frac{{\partial \left| {\Psi \left( t \right)} \right\rangle }}{{\partial t}} = H\left| {\Psi \left( t \right)}
 \right\rangle.
 \end{equation}

 Since the electron-pulse dipole interaction gives rise to the one-particle excitations only,
 we cannot generate an entanglement in a two-electron system
 using the one-photon process only. However, it is seen from the energy level structure
 that the states $\left| 1 \right\rangle $ and $\left| 2 \right\rangle $
 can be coupled by the four-stage two-photon transition scheme involving the ground
states, the states from the subspace with the energy
$\varepsilon_{01}^{\pm}$, and the hybridized $\left| 13
\right\rangle $ - $\left| 16 \right\rangle $ or doubly-occupied
$\left| 3 \right\rangle $, $\left| 4 \right\rangle $ states.

  First, we consider the transition scheme that involves the states $\left| 1 \right\rangle $,
   $\left| 2 \right\rangle $, $\left| 5 \right\rangle $ - $\left| 8 \right\rangle $, and $\left|
13 \right\rangle $ (Fig. 2). Because of the degeneracy of the
states $\left| 5 \right\rangle $ - $\left| 8 \right\rangle $, two
pulses are sufficient to induce the transitions between the states
$\left| 1 \right\rangle $ and $\left| 2 \right\rangle $. We set
$\omega _1  = \varepsilon _{01}^ - - \varepsilon _{00} + \Delta _1
$ and $\omega _2  = \varepsilon _{11}^{- s }  - \varepsilon _{01}^
-   + \Delta _2 $, where $\Delta _1 $ and $\Delta _2 $ are the
detunings. The transformation of the coefficients $c_n$ in Eq. (6)
 according to $c_1  = \tilde c_1 ,\,\,c_2
= \tilde c_2 $, $c_k  = \tilde c_k \exp \left( { - i\Delta _1 t}
\right)$ for $k = 5 - 8$, $c_{13}  = \tilde c_{13} \exp \left[ { -
i\left( {\Delta _1  + \Delta _2 } \right)t} \right]$ and their
substitution into Eqs. (6) and (7) result in the set of equations:
\begin{equation}
 \left\{ \begin{array}{l}
 i\dot {\tilde c_1}  = \lambda _1 \left( {\tilde c_5  + \tilde c_6 } \right), \\
 i\dot {\tilde c_2}  = \lambda _1 \left( {\tilde c_7  + \tilde c_8 } \right), \\
 i\dot {\tilde c_k}  =  - \Delta _1 \tilde c_k  + \lambda _1^* \left( {\delta _{k 5} \tilde c_1 +\delta _{k 6} \tilde c_1 + \delta _{k 7} \tilde c_2 +\delta _{k 8} \tilde c_2} \right) + \lambda _2 \tilde c_{13} ,\,\,\,k = 5 - 8 \\
 i\dot {\tilde c_{13}}  =  - \left( {\Delta _1  + \Delta _2 } \right)\tilde c_{13}  + \lambda _2^* \left( {\tilde c_5  + \tilde c_6  + \tilde c_7  + \tilde c_8 } \right), \\
 \end{array} \right.
\end{equation}
where $\lambda _1  = \frac{1}{2}\exp \left( {i\varphi _1 }
\right)\left\langle 1 \right|V_1 \left| 5 \right\rangle $ and
$\lambda _2  = \frac{1}{2}\exp \left( {i\varphi _2 }
\right)\left\langle 5 \right|V_2 \left| {13} \right\rangle $ are
the matrix elements of the electron-pulse interaction (here we
take into account the identities $\left\langle 1 \right|V_1 \left|
5 \right\rangle  = \left\langle 1 \right|V_1 \left| 6
\right\rangle =
 \left\langle 2 \right|V_1 \left| 7 \right\rangle  = \left\langle 2 \right|V_1 \left| 8 \right\rangle $ and
$\left\langle 5 \right|V_2 \left| {13} \right\rangle  =
\left\langle 6 \right|V_2 \left| {13} \right\rangle  =
\left\langle 7 \right|V_2 \left| {13} \right\rangle
 = \left\langle 8 \right|V_2 \left| {13} \right\rangle $).

The eigenfrequencies of the set of Eqs. (8) can be found from the
following equation:
\begin{equation}
 x^3  - i\left( {2\Delta _1  + \Delta _2 } \right)x^2  + \left[ {4\left| {\lambda _2 } \right|^2  + 2\left| {\lambda _1 } \right|^2  - \Delta _1 \left( {\Delta _1  + \Delta _2 } \right)}
 \right]x  - 2i\left| {\lambda _1 } \right|^2 \left( {\Delta _1  + \Delta _2 } \right) = 0.
\end{equation}
The solution of Eq. (9) is straightforward. However, only in the
case that $\Delta _1  =  - \Delta _2  = \Delta $ it may be
presented in a rather simple form. This case corresponds to the
two-photon resonance.

  If the system is initially in the superposition of the states $\left| 1 \right\rangle $
and $\left| 2 \right\rangle $, i. e., $\left| {\Psi \left( 0
\right)} \right\rangle  = \alpha \left| 1 \right\rangle  + \beta
\left| 2 \right\rangle $, the initial conditions are $\tilde{c_1}
\left( 0 \right) = \alpha ,\,\,\tilde{c_2} \left( 0 \right) =
\beta ,\,\,\tilde{c}_{k \ne 1,2} \left( 0 \right) = 0$ and we
obtain from the set of Eqs. (8) the following expressions for the
coefficients $c_n $ in the laboratory frame:
\begin{equation}
 \begin{array}{l}
 c_{1,2}  =  \pm \frac{{{\rm{\alpha }} - {\rm{\beta }}}}{2}\exp \left( {i\frac{{\Delta t}}{2}} \right)\left[ {\cos \left( {\Omega _1 t} \right) - i\frac{\Delta }{{2\Omega _1 }}\sin \left( {\Omega _1 t} \right)} \right] +  \\
 \,\,\,\,\,\,\,\,\,\,\,\, + \frac{{{\rm{\alpha }} + {\rm{\beta }}}}{2}\left\{ {\frac{{2\left| {\lambda _2 } \right|^2 }}{{\left| {\lambda _1 } \right|^2  + 2\left| {\lambda _2 } \right|^2 }} + \exp \left( {i\frac{{\Delta t}}{2}} \right)\frac{{\left| {\lambda _1 }
 \right|^2 }}{{\left| {\lambda _1 } \right|^2  + 2\left| {\lambda _2 } \right|^2 }}\left[ {\cos \left( {\Omega _2 t} \right) - i\frac{\Delta }{{2\Omega _2 }}\sin \left( {\Omega _2 t} \right)} \right]} \right\} \\
 \end{array},
\end{equation}
\begin{equation}
 c_5  = c_6 = - i\exp \left( {i\frac{{\Delta t}}{2}} \right)\left[ { \frac{{\alpha  - \beta }}{2}\frac{{\lambda _1^* }}{{\Omega _1 }}\sin \left( {\Omega _1 t} \right) +
 \frac{{\alpha  + \beta }}{2}\frac{{\lambda _1^* }}{{\Omega _2 }}\sin \left( {\Omega _2 t} \right)}
 \right],
\end{equation}
\begin{equation}
 c_7  = c_8 = - i\exp \left( {i\frac{{\Delta t}}{2}} \right)\left[ { -\frac{{\alpha  - \beta }}{2}\frac{{\lambda _1^* }}{{\Omega _1 }}\sin \left( {\Omega _1 t} \right) +
 \frac{{\alpha  + \beta }}{2}\frac{{\lambda _1^* }}{{\Omega _2 }}\sin \left( {\Omega _2 t} \right)}
 \right],
\end{equation}
\begin{equation}
 c_{13}  = \left( {\alpha  + \beta } \right)\frac{{\lambda _1^* \lambda _2^* }}{{\left| {\lambda _1 } \right|^2  + 2\left| {\lambda _2 } \right|^2 }}\left\{ { - 1 + \exp \left( {i\frac{{\Delta t}}{2}} \right)\left
[ {\cos \left( {\Omega _2 t} \right) - \frac{{i\Delta }}{{2\Omega
_2 }}\sin \left( {\Omega _2 t} \right)} \right]} \right\}.
\end{equation}
Here $\Omega _1  = \sqrt {\frac{{\Delta ^2 }}{4} + 2\left|
{\lambda _1 } \right|^2 } $ and $\Omega _2  = \sqrt {\frac{{\Delta
^2 }}{4} + 2\left( {\left| {\lambda _1 } \right|^2 + 2\left|
{\lambda _2 } \right|^2 } \right)} $ are the Rabi frequencies.

 If $\Delta  = 0$, the system is in the exact resonance with both pulses. To localize the system completely in the ground-state subspace $\left\{\left| 1 \right\rangle, \left| 2 \right\rangle \right\}  $,
 the matrix elements of the electron-pulse interaction ${\lambda _1 }$  and ${\lambda _2 }$ must satisfy  the condition
\begin{equation}
 \frac{{\left| {\lambda _2 } \right|}}{{\left| {\lambda _1 } \right|}} = \sqrt {\frac{1}{2}\left( {\frac{{4m^2 }}{{{k}^2 }} - 1} \right)} ,
\end{equation}
where $k,\,m$ are integers chosen so that the right-hand side of
Eq. (14) be real. The Eq. (14) ensures that $c_{n \ne 1,2} \left(
{T_k } \right) = 0$. We see that in this case for the pulse
durations $T_k  = {{\pi k} \mathord{\left/
 {\vphantom {{\pi k} {\left| {\lambda _1 } \right|\sqrt 2 }}} \right.
 \kern-\nulldelimiterspace} {\left| {\lambda _1 } \right|\sqrt 2 }}$
the inversion operation ${\rm{\sigma }}_x $ is realized in the
ground-state subspace if $k$ is odd, while the initial state is
not changed if $k$ is even.

 There is an interesting particular case of
$\Delta  = {{\left( {\varepsilon _{11}^{ - s}  + \varepsilon _{00}
} \right)} \mathord{\left/
 {\vphantom {{\left( {\varepsilon _{11}^{ - s}  + \varepsilon _{00} } \right)} 2}} \right.
 \kern-\nulldelimiterspace} 2} - \varepsilon _{01}^{-} $ similar to that discussed in Ref. \cite{28}.
 In this case $\omega _1  = \omega _2 $ and only one pulse is
sufficient for the state inversion. If the value of $\Delta $ is
much larger than the matrix elements of the electron-pulse
interaction, the populations $\left| {c_k } \right|^2$  of the
auxiliary states $\left| 5 \right\rangle $ - $\left| 8
\right\rangle $ are of the order of $\left( {{{\left| {\lambda _1
} \right|} \mathord{\left/
 {\vphantom {{\left| {\lambda _1 } \right|} \Delta }} \right.
 \kern-\nulldelimiterspace} \Delta }} \right)^2  <  < 1$, i. e.,
 much smaller than the populations of states $\left| 1 \right\rangle
$, $\left| 2 \right\rangle $ and $\left| 13 \right\rangle $. It
corresponds to the well-known adiabatic elimination procedure,
widely used in the quantum optics (see, e. g., Ref. \cite{45}).
The auxiliary states $\left| k \right\rangle $ with $k = 5 - 8$
are populated only virtually, and the effective three-level scheme
is realized, giving rise to the simultaneous excitation of two
electrons from the ground-state subspace to the state $\left| 13
\right\rangle $. We may view this process as the generation of the
effective exchange coupling between the electron spins driven by
one resonant pulse. The corresponding dynamics can be obtained
through the expansion Eqs. (10) - (13) in terms of the small
parameters ${{\left| {\lambda _{1,2} } \right| } \mathord{\left/
 {\vphantom {{\left| {\lambda _{1,2} } \right| } {\left| \Delta  \right|}}} \right.
 \kern-\nulldelimiterspace} {\left| \Delta  \right|}} <  < 1$. This process is slow compared to that with
$\Delta=0$, since its Rabi frequency
 is $\Omega _{eff}  = {{\left| {\lambda _1 } \right|^2 }
\mathord{\left/
 {\vphantom {{\left| {\lambda _1 } \right|^2 } {\left| \Delta  \right|}}} \right.
 \kern-\nulldelimiterspace} {\left| \Delta  \right|}} <  < \left| {\lambda _1 }
 \right|$.
 The stroboscopical evolution of the state vector in the ground state
 subspace $\left| {\Psi \left( {T_k } \right)} \right\rangle  =
\exp \left( { - i\varepsilon _{00} T_k } \right)U_k \left| {\Psi
\left( 0 \right)} \right\rangle $, where $\left| {\Psi \left( 0
\right)} \right\rangle = \left(\alpha , \beta \right) ^{T}$ and
$T_k  = {{\pi k\Delta } \mathord{\left/
 {\vphantom {{\pi k\left| {\lambda _1 } \right|^2 } {\left( {\left| {\lambda _1 } \right|^2  + 2\left| {\lambda _2 } \right|^2 } \right)}}} \right.
 \kern-\nulldelimiterspace} {\left( {\left| {\lambda _1 } \right|^2  + 2\left| {\lambda _2 } \right|^2 } \right)}}$, is given by the matrix $U_{k}$:
\begin{equation}
 U_{k} = \exp \left( { - i\psi _k } \right)\left( {\begin{array}{*{20}c}
   {\cos \left( {\psi _k } \right)} & {\exp \left( {i{\pi  \mathord{\left/
 {\vphantom {\pi  2}} \right.
 \kern-\nulldelimiterspace} 2}} \right)\sin \left( {\psi _k } \right)}  \\
   { - \exp \left( { - i{\pi  \mathord{\left/
 {\vphantom {\pi  2}} \right.
 \kern-\nulldelimiterspace} 2}} \right)\sin \left( {\psi _k } \right)} & {\cos \left( {\psi _k } \right)}  \\
\end{array}} \right)
,
\end{equation}
where $\psi _k  = \Omega_{eff} T_k ={{\pi k \left| {\lambda _1 }
\right|^2} \mathord{\left/
 {\vphantom {{\pi k\left| {\lambda _1 } \right|^2 } {\left( {\left| {\lambda _1 } \right|^2  + 2\left| {\lambda _2 } \right|^2 } \right)}}} \right.
 \kern-\nulldelimiterspace} {\left( {\left| {\lambda _1 } \right|^2  + 2\left| {\lambda _2 } \right|^2 } \right)}}$. Here the
pulse duration $T_k$ is determined by the condition $c_{13} \left(
{T_k } \right) = 0$. We see that Eq. (15) corresponds to the
in-plane rotation through the angle $\psi _k $. This operation
creates the maximally entangled state for the single-spin qubit
encoding scheme of Ref. \cite{9}
 but, of course, is not sufficient for an arbitrary rotation of the qubit state vector on the Bloch sphere for the scheme where the states
$\left| 1 \right\rangle $ and $\left| 2 \right\rangle $ are used
as qubit states (see below).

 Next we consider the situation where $\Delta _1  \ne \Delta _2 $ and one (or both) of the transitions is off-resonant.
Assuming $\Delta _1  = 0,\,\,\,\left| {\Delta _2 } \right| >  >
\left| {\lambda _2 } \right|$, we exclude adiabatically the
singlet state from the set of Eqs. (8) as follows:
\begin{equation}
 \dot {\tilde c_{13}}  = 0,\,\,\, \tilde c_{13}  = \frac{{\lambda _2^*
}}{{\Delta _2 }}\left( {\tilde c_5  + \tilde c_6  + \tilde c_7  +
\tilde c_8 } \right),
\end{equation}
and substitute Eq. (16) into Eq. (8), thus arriving at the
following expressions:
\begin{equation}
 c_{1,2}  =  \pm \frac{{\alpha  - \beta }}{2}\cos \left( {\sqrt 2 \left| {\lambda _1 } \right|t} \right) + \frac{{\alpha  + \beta }}{2}\exp \left( { - i\frac{2{\left| {\lambda _2 } \right|^2 t}}{{\Delta _2 }}} \right)\left
[ {\cos \left( {\tilde \Omega _2 t} \right) + i\frac{2{\left|
{\lambda _2 } \right|^2 }}{{\Delta _2 \tilde \Omega _2 }}\sin
\left( {\tilde \Omega _2 t} \right)} \right],
\end{equation}
\begin{equation}
c_5  = c_6 = - \frac{i}{{\sqrt 2 }}\left[ {\frac{{\alpha  - \beta
}}{2}\sin \left( {\sqrt 2 \left| {\lambda _1 } \right|t} \right) +
\frac{{\alpha  + \beta }}{2}\frac{{\sqrt 2 \lambda _1^* }}{{\tilde
\Omega _2 }}\exp \left ( { - i\frac{{2\left| {\lambda _2 }
\right|^2 t}}{{\Delta _2 }}} \right)\sin \left( {\tilde \Omega _2
t} \right)} \right],
\end{equation}
\begin{equation}
c_7  = c_8 = - \frac{i}{{\sqrt 2 }}\left[ { -\frac{{\alpha  -
\beta }}{2}\sin \left( {\sqrt 2 \left| {\lambda _1 } \right|t}
\right) + \frac{{\alpha  + \beta }}{2}\frac{{\sqrt 2 \lambda _1^*
}}{{\tilde \Omega _2 }}\exp \left ( { - i\frac{{2\left| {\lambda
_2 } \right|^2 t}}{{\Delta _2 }}} \right)\sin \left( {\tilde
\Omega _2 t} \right)} \right],
\end{equation}

where $\tilde \Omega _2  = \sqrt {\frac{4{\left| {\lambda _2 }
\right|^4 }}{{\Delta _2^2 }} + 2\left| {\lambda _1 } \right|^2 }
$. As for the cases discussed above, there is a relation between
the matrix elements of the electron-pulse interaction necessary to
achieve a complete localization of the system in the subspace
spanned by the states $\left| 1 \right\rangle $ and $\left| 2
\right\rangle$:
\begin{equation}
\frac{{\left| {\lambda _2 } \right|}}{{\left| {\lambda _1 }
\right|}} = \frac{{\left| {\Delta _2 } \right|}}{{\sqrt 2  \left|
{\lambda _2 } \right|}}\sqrt {\frac{{m^2 }}{{k^2 }} - 1} ,
\end{equation}
where $m,\,k\,$ are integers. If this condition is fulfilled, the
ground state evolution matrix at $T_k  = {{\pi k} \mathord{\left/
 {\vphantom {{\pi k} {\left| {\lambda _1 } \right|\sqrt 2 }}} \right.
 \kern-\nulldelimiterspace} {\left| {\lambda _1 } \right|\sqrt 2 }}$
  is given by the expression:
\begin{equation}
 U_{k} = \exp \left( i\pi k \right) \exp \left( { - i\psi _k } \right)\left( {\begin{array}{*{20}c}
   {\cos \left( {\psi _k } \right)} & {\exp \left( {-i{\pi  \mathord{\left/
 {\vphantom {\pi  2}} \right.
 \kern-\nulldelimiterspace} 2}} \right)\sin \left( {\psi _k } \right)}  \\
   { - \exp \left( { i{\pi  \mathord{\left/
 {\vphantom {\pi  2}} \right.
 \kern-\nulldelimiterspace} 2}} \right)\sin \left( {\psi _k } \right)} & {\cos \left( {\psi _k } \right)}  \\
\end{array}} \right)
,
\end{equation}
 with $\Omega _{eff}  = {{\left| {\lambda _2 } \right|^2 }
\mathord{\left/
 {\vphantom {{\left| {\lambda _2 } \right|^2 } {\left| {\Delta _2 } \right|}}} \right.
 \kern-\nulldelimiterspace} {\left| {\Delta _2 } \right|}}$, $\psi _k  = \Omega_{eff} T_k = \pi\sqrt{m^2-k^2}/2$ if $m-k$ is even and $\tilde U_k = \exp \left( {i\pi }
\right){\rm{\sigma }}_x U_{k} = \exp \left( {i\pi } \right)U_{k}
{\rm{\sigma }}_x $ if $m-k$ is odd.

 The cases $\Delta _2  = 0,\,\left| {\Delta _1 } \right| >  > \left| {\lambda _1 } \right|$ and
$\left| {\Delta _1 } \right|,\,\left| {\Delta _2 } \right| >  >
\left| {\lambda _1 } \right|,\,\left| {\lambda _2 } \right|$
enable a complete localization of the system in the ground state
subspace at any time without  restrictions on the matrix elements
of the electron-pulse interaction. However, the former case does
not reveal the non-trivial evolution up to the fourth order in the
ratio ${{\left| {\lambda _1 } \right|} \mathord{\left/
 {\vphantom {{\left| {\lambda _1 } \right|} {{\left| \Delta _1 \right|}}}} \right.
 \kern-\nulldelimiterspace} {\Delta _1 }}$, and
 the effective Rabi
frequency in the latter case is $\Omega _{eff}  \sim {{\left|
{\lambda _1 } \right|^2 \left| {\lambda _2 } \right|^2 }
\mathord{\left/
 {\vphantom {{\left| {\lambda _1 } \right|^2 \left| {\lambda _2 } \right|^2 } {\Delta _1^2 \left| {\Delta _2 } \right|}}} \right.
 \kern-\nulldelimiterspace} {\Delta _1^2 \left| {\Delta _2 } \right|}}$
 that makes the rotations defined by
Eq. (15) too slow and unviable in view of spin decoherence
processes. Here we don't consider these cases in details.

 Next we study another way to achieve a rotation of the quantum state of the two-electron system. As it was mentioned before, the doubly-occupied states
$\left| 3 \right\rangle $ and $\left| 4 \right\rangle $ may also
be exploited for the creation of the effective exchange coupling
between the electrons localized in $\left| 1 \right\rangle $ or
$\left| 2 \right\rangle $ states. To see how it may be realized we
present the scheme for the electron transitions that provides such
coupling (Fig. 3). We see that only the states $\left| 1
\right\rangle $ - $\left| 8 \right\rangle $ are involved in the
process, while the doubly-excited states $\left| 13 \right\rangle
$ - $\left| 16 \right\rangle $ do not participate the dynamics. In
complete analogy with the procedure described above, we represent
the state vector of the system in terms of the stationary
eigenstates of Eq. (1) with the time-dependent coefficients $c_k
$, where $k=1 - 8$. Next we substitute it into Eq. (7), where the
frequencies of two laser pulses now are given by the expressions
$\omega _1  = \varepsilon _{01}^ -   - \varepsilon _{00}  + \Delta
_1 $ and $\omega _2  = \varepsilon _{01}^ -   - \tilde \varepsilon
_{00}^{} - \Delta _2 $. The coefficients may be obtained from the
set of equations:
\begin{equation}
 \left\{ \begin{array}{l}
 i\dot {\tilde c_1}  = \lambda _1 \left( {\tilde c_5  + \tilde c_6 } \right), \\
 i\dot {\tilde c_2}  = \lambda _1 \left( {\tilde c_7  + \tilde c_8 } \right), \\
 i\dot {\tilde c_3}  =  - \left( {\Delta _1  + \Delta _2 } \right)\tilde c_3  + \tilde\lambda _2^{*} \left( {\tilde c_5  + \tilde c_8 } \right), \\
 i\dot {\tilde c_4}  =  - \left( {\Delta _1  + \Delta _2 } \right)\tilde c_4  + \tilde\lambda _2^{*} \left( {\tilde c_6  + \tilde c_7 } \right), \\
 i\dot {\tilde c_k}  =  - \Delta _1 \tilde c_k  + \lambda _1^* \left( {\delta _{k5} \tilde c_1 +\delta _{k6} \tilde c_1 + \delta _{k7} \tilde c_2 +\delta _{k8} \tilde c_2} \right) \\
  + \tilde\lambda _2 \left( {\delta _{k5} \tilde c_3 +\delta _{k8} \tilde c_3 + \delta _{k6} \tilde c_4 +\delta _{k7} \tilde c_4} \right),\,\,\,k = 5 - 8, \\
 \end{array} \right.
\end{equation}
where now $\tilde\lambda _2  = \frac{1}{2}\exp \left( {i\varphi _2
} \right)\left\langle 5 \right|V_1 \left| 3 \right\rangle $ and
the identities $\left\langle 5 \right|V_2 \left| 3 \right\rangle =
\left\langle 8 \right|V_2 \left| 3 \right\rangle  = \left\langle 6
\right|V_2 \left| 4 \right\rangle  =
 \left\langle 7 \right|V_2 \left| 4 \right\rangle $ are assumed.

 The set of Eqs. (22) is equivalent to the set of Eqs. (8) if we
 set in Eqs. (22) $\tilde c_3=\tilde c_4={\tilde c_{13}}/\sqrt{2}$ and ${\tilde \lambda_2}=\lambda_2{\sqrt{2}}
 $. This substitution allowed in the case $c_{13}(0)=0, c_{3}(0)=0, c_{4}(0)=0$ provides
 the formal analogy between the first and the
second schemes. Thus one can derive all time-dependent
coefficients $c_k$ relevant for the quantum dynamics from Eqs.
(10) - (21).

 Note that apart from the restrictions imposed by the state localization
 requirement, the matrix elements of the electron-pulse interaction, as well as the detunings
 from resonance, have to satisfy another condition, justifying the
 resonant approximation used in Eqs. (8) and Eqs. (22). If we
 consider the transitions involving the states $\left| 5
\right\rangle $ - $\left| 8 \right\rangle $ or the states $\left|
9 \right\rangle $ - $\left| 12 \right\rangle $, the following
inequalities must be fulfilled: $\left| {\Delta _1 }
\right|,\,\left| {\Delta _2 } \right| <  < \left| {\lambda _1 }
\right|,\,\left| {\lambda _2 } \right| <  < \varepsilon _{01}^ + -
\varepsilon _{01}^ -  $. For the off-resonant transition we should
keep $\left| {\Delta } \right|<<\varepsilon _{01}^ +   -
\varepsilon _{01}^ -  $. (Of course, the energy differences of the
states $\left| 13 \right\rangle $ - $\left| 16 \right\rangle $
must satisfy similar inequalities as well.) If we take the matrix
elements of the electron-pulse interaction $\left| {\lambda
_{1,\,2} } \right| \sim 10^{ - 5} $ eV and assume that $\tilde t
\sim 10^{ - 3} $ eV and $\tilde t \sim \varepsilon _{01}^ + -
\varepsilon _{01}^ -  $, it is possible to meet the condition
above in both resonant and off-resonant cases. In the resonant
case, this choice of the structure parameters implies the
operation times $\tau \sim {1} / {{\rm min} \left( \left| {\lambda
_{1}} \right|, \left| {\lambda _{2} } \right| \right|)}$ to be of
the order of tens of picoseconds. The setting of the matrix
elements of the electron-pulse interaction $\lambda _{1,\,2}$ can
be readily achieved via adjusting the strengths of the pulses.

   In the next section, we consider several important potential applications of the
  results obtained to the quantum information processing.

\vskip 6mm

\centerline{\bf III. ENTANGLEMENT AND QUANTUM STATE ENGINEERING}

Electron spin localized in the QD is now extensively
 studied as the promising candidate for the solid-state qubit implementation.
Let us investigate the possibility of exploiting the electron
spins in the double QD structure for quantum computations in view
of the realization of quantum operations by optical means. We
discuss the situation where the states $\left| 1 \right\rangle =
\left| {1_ \uparrow  ,1_ \downarrow ,0,0} \right\rangle $ and
$\left| 2 \right\rangle  = \left| {1_ \downarrow  ,1_ \uparrow
,0,0} \right\rangle $ are used as the qubit logical states
\cite{46}. Thus, a double-dot structure is now viewed as a single
qubit. What kind of quantum operations may be performed on such a
qubit? We have found from Eqs. (15) and (21) that the rotation
along the meridian $\varphi  = {\pi \mathord{\left/ {\vphantom
{\pi  2}} \right.
 \kern-\nulldelimiterspace} 2}$ through the polar angle $\psi_k$ may be realized.
 It is easy to see, however, that the states
 $\left| 1 \right\rangle $ and $\left| 2 \right\rangle $ belong to the same charge
configuration. To generate, say, $\sigma_z$ operation one should
distinguish between them, making use of the Pauli exclusion
principle only. The charge and spin degrees of freedom are
decoupled from each other, and the laser pulse cannot change the
electron spin in a given QD.  To construct a quantum gate that
acts directly on the electron charge only but is able to operate
with the spin-encoded qubit, some additional tools are required.
For example, one can exploit the exciton-based techniques
\cite{32}, where the circularly-polarized laser pulse generates
the electron transition to the conduction band depending on the
spin of the electron occupying the lowest size-quantized level.

 With the help of the optics it is possible to transport an individual spin through the
sequence of the QDs by swapping electron spins in the neighboring
QDs, induced by the laser pulse that generates $\sigma_x$
operation. This effect may be helpful for the quantum state
transfer in the models based on the Loss - DiVincenzo's proposal.

   There exist several more spin qubit encoding schemes \cite{47} - \cite{50}. We can
arrive at these schemes  by organizing the entanglement between
the QD spins (see below), that is equivalent to the superposition
of the states $\left| 1 \right\rangle $  and $\left| 2
\right\rangle $, and using the entangled states as logical ones.
This offers an opportunity to handle with quantum information
within the decoherence-free subspaces that, in its turn, provides
robust quantum information processing. Note that in the four-spin
encoding scheme of Ref. \cite{48} both $\sigma_x$ and $\sigma_z$
operations can be performed via the sequence of the corresponding
two-spin inversion operations.

 Next we discuss the possible application of the optically-induced
quantum evolution of the two-electron system in the entanglement
generation. A lot of proposals based on a few-electron QD, using
the static and/or alternating electric field controlling
techniques, have been made  to achieve this purpose \cite{35} -
\cite{44}. Some of them seem to be very promising, especially
those handling with two interacting electrons in the double QD
under the action of an oscillatory electric field. As it was
demonstrated, both static \cite{40} - \cite{44} and alternating
 \cite{35} - \cite{39} electric fields satisfying some conditions
may be used to entangle two electrons and localize them in the
entangled state. Besides, by an appropriate switching procedure
one can drive the system between the delocalized and fully
localized states and coherently destroy the electron tunneling
process. In those schemes, the orbital electron states localized
in the left (right) QD serve as the logical states. The Coulomb
repulsion plays a significant role in such systems; depending on
the Hubbard ratio $t/U$, several regimes may be realized. All of
the dynamical properties can be derived through the analysis of
the Floquet spectrum of the two-electron system in the electric
field.

 We present one another way to produce an entanglement in the two-electron double-dot structures.
If we treat the states  $\left| 1 \right\rangle  = \left| {1_
\uparrow  ,1_ \downarrow  ,0,0} \right\rangle $ and $\left| 2
\right\rangle  = \left| {1_ \downarrow  ,1_ \uparrow ,0,0}
\right\rangle $ as the two-qubit states in the Loss - DiVincenzo's
scheme, it is easy to create the maximally entangled Bell state of
two spins ${{\left( {\left| 1 \right\rangle  + i\left| 2
\right\rangle } \right)} \mathord{\left/
 {\vphantom {{\left( {\left| 1 \right\rangle  - i\left| 2 \right\rangle } \right)} {\sqrt 2 }}} \right.
 \kern-\nulldelimiterspace} {\sqrt 2 }}$ as follows. One may use the single pulse with
 the detuning $\Delta  = {{\left( {\varepsilon _{11}^{ - s}  + \varepsilon _{00}
} \right)} \mathord{\left/
 {\vphantom {{\left( {\varepsilon _{11}^{ - s}  + \varepsilon _{00} } \right)} 2}} \right.
 \kern-\nulldelimiterspace} 2} - \varepsilon _{01}^{-} $ and
$\left| {\lambda _2 } \right| = \sqrt {{3 \mathord{\left/
 {\vphantom {3 2}} \right.
 \kern-\nulldelimiterspace} 2}} \left| {\lambda _1 } \right|$.
 The evolution described by Eq.
(15) is then realized, and if the system has started from the
state $\left| 1 \right\rangle$, the Bell state (up to the common
phase) is obtained at $\tau _{Bell}  = {{\pi \left| {\Delta }
\right|} \mathord{\left/
 {\vphantom {{\pi \left| {\Delta } \right|} 4{\left| {\lambda _1 } \right|}}} \right.
 \kern-\nulldelimiterspace} 4{\left| {\lambda _1 } \right|}}^2 $.
  Note that this state may be realized in two ways
considered in Sec. II B. Besides of the formation of the spin
entangled states, there is also a possibility to generate the
entanglement between the charge states, as it was introduced in
Refs. \cite{35}, \cite{39}. To do this one should start from the
superposition of the states like ${{\left( {\left| 1 \right\rangle
+ \left| 2 \right\rangle } \right)} \mathord{\left/
 {\vphantom {{\left( {\left| 1 \right\rangle  + \left| 2 \right\rangle } \right)} {\sqrt 2 }}} \right.
 \kern-\nulldelimiterspace} {\sqrt 2 }}$ or from one of the doubly-occupied
 states. The entangled states of two different forms, $\left(\left| 2,0,0,0 \right\rangle \pm   \left| 0,2,0,0 \right\rangle \right) /{\sqrt 2 }$ and
 $\left(\left| 0,0,2,0 \right\rangle \pm   \left| 0,0,0,2 \right\rangle \right) /{\sqrt 2 }$, can be obtained
 by the use of the corresponding transition scheme.
If one needs to obtain the state similar to that discussed in Ref.
\cite{39}, i. e. $\left(\left| 0,0,2,0 \right\rangle +  \left|
0,0,0,2 \right\rangle \right) /{\sqrt 2 }$, one should replace the
state $\left| 13 \right\rangle $ by the state $\left| 16
\right\rangle $ in the transition scheme of Eqs. (10) - (13) and
tune the system parameters so that the regime with $\tilde t <<
\tilde v$ be realized. Then the state $\left| 16 \right\rangle $
turns out to be the equally-weighted superposition of the
doubly-occupied excited states. Of course, the quantum state
engineering procedure
 retains all of the operation tools, e. g., Coulomb repulsion control proposed earlier \cite{35}. On the other hand, the entanglement of the doubly-occupied states
  $\left| 3 \right\rangle  $ and $\left| 4 \right\rangle $ may be achieved in the transition scheme presented by Eqs. (22) without any
modification.

  Here we draw attention to the main distinguishing features of our model from those mentioned above. First,
we consider the two-level QDs instead of the single-level QDs
studied in Refs. \cite{35} - \cite{44}. This enables us to
localize an electron in the well-isolated ground-state subspace
without use of any additional resource. Second, we operate with
the pulse intensities much smaller as compared to those used in
the cited works. It would help one to carefully isolate the
logical and auxiliary states from other ones during the pulse
action. The resonant character of the structure-field interaction
serves for the same purpose. What we would like to mention, our
scheme turns out to be useful in the entanglement generation for
both spin and charge encoding schemes.

\vskip 5mm

\centerline{\bf IV. CONCLUSIONS}

We have considered  the quantum dynamics of two interacting
electrons confined in the double-dot structure. The quantum
transitions occur between the eigenstates of the stationary
two-electron Hamiltonian under influence of the resonant laser
pulse. By an appropriate choice of the pulse parameters we may
generate the superposition of some states. There are several ways
for the quantum state manipulations by the use of the
electromagnetic field. All of them involve the hybridized or
doubly-occupied states as the auxiliary states. Those states play
an important role in the non-trivial quantum state evolution and
serve as "the entanglers" in the schemes presented here. The
maximally-entangled Bell states may be generated for both electron
spin states and electron charge states. We can say that the laser
pulse generates (and destroys) the effective exchange coupling
between the electron spins.
 This makes the two-electron double-dot
system very interesting for the quantum information processing.
Several important quantum operations may be constructed using this
system. If we consider the doubly-degenerate ground states (or
their combinations) as the logical states, the optically-driven
single qubit operation $\sigma_x$ may be realized in the logical
subspace.
 In this work we have studied a simplified model to catch the
principal features of the behavior of two confined interacting
electrons in the resonant field. The next steps of investigations
are the study of the influence of structure asymmetry and the
pulse imperfections on the quantum dynamics as well as the
decoherence effects. However, one should expect that further
analysis will retain all of qualitative results obtained here. We
hope that our study will be helpful for the solid-state quantum
computer design.

\vskip 4mm

\centerline{\bf ACKNOWLEDGMENTS}

Author would like to thank L.A. Openov for the critical reading of
the manuscript and K.A. Valiev for his interest in this study.

\newpage

\includegraphics[width=\hsize]{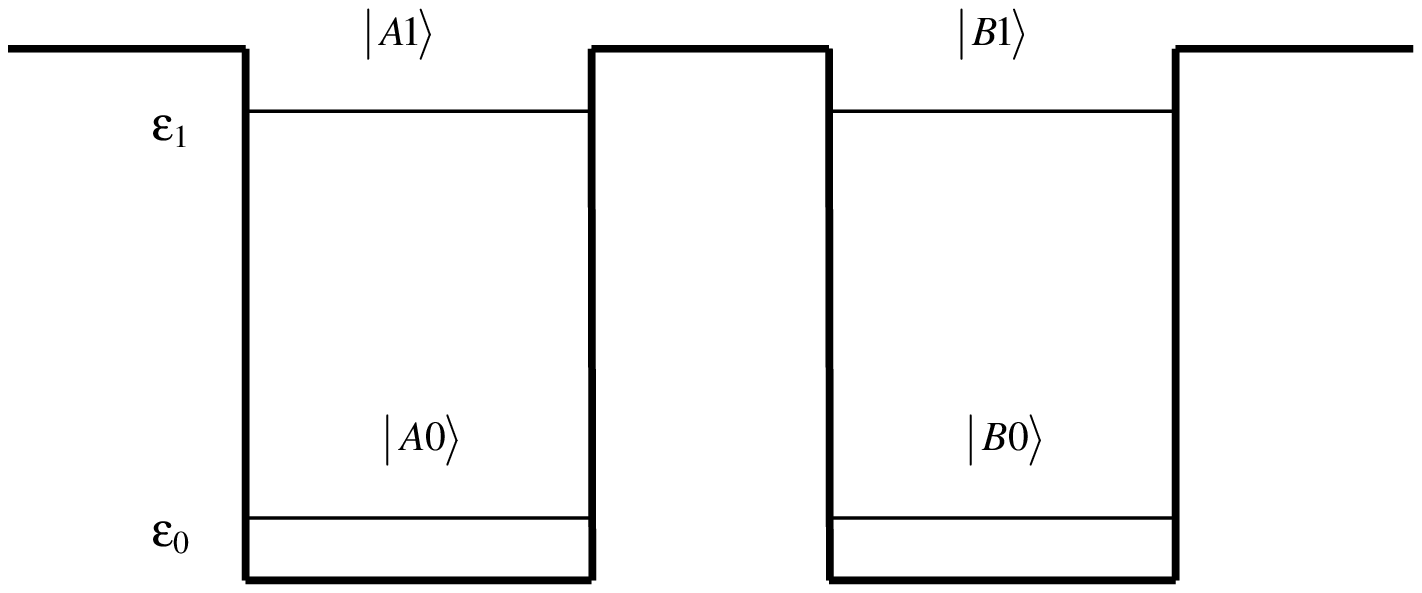}

\vskip 6mm

Fig. 1. Schematics of the states for a single electron confined in
the double-dot structure. These one-electron states are used for
construction of the two-electron states, see text for details.

\newpage

\includegraphics[width=\hsize]{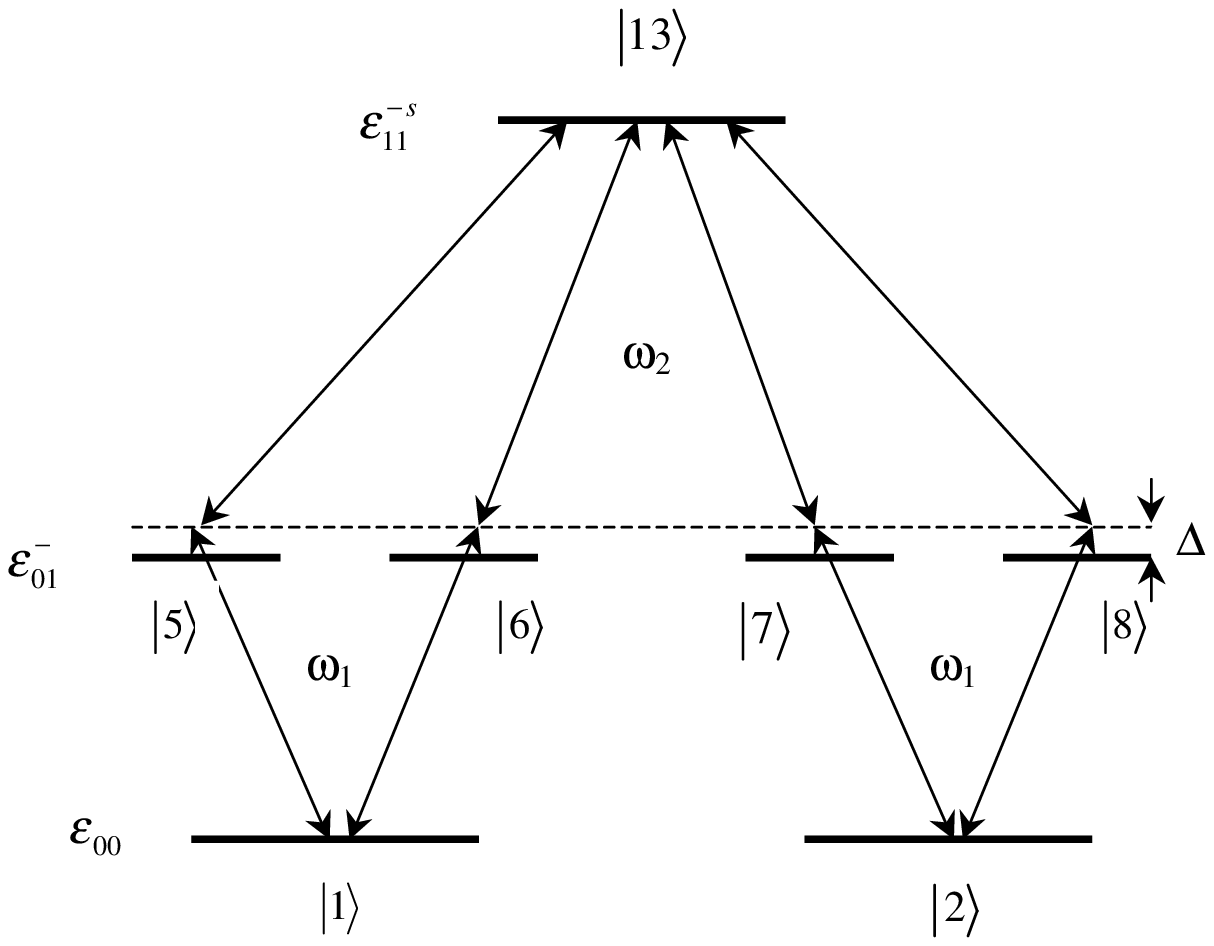}

\vskip 6mm

Fig. 2. Transition scheme connecting the states $\left| 1
\right\rangle $ and $\left| 2 \right\rangle $ through the use of
the auxiliary states $\left| 5 \right\rangle $, $\left| 6
\right\rangle $, $\left| 7 \right\rangle $, $\left| 8\right\rangle
$, and $\left| 13 \right\rangle $ in the case of the two-photon
resonance $\Delta_1=-\Delta_2=\Delta$. Here $\Delta_1$ and $
\Delta_2$ are the detunings of the pulse frequencies $\omega_1$
and $\omega_2$, respectively, from the exact resonance.

\newpage

\includegraphics[width=\hsize]{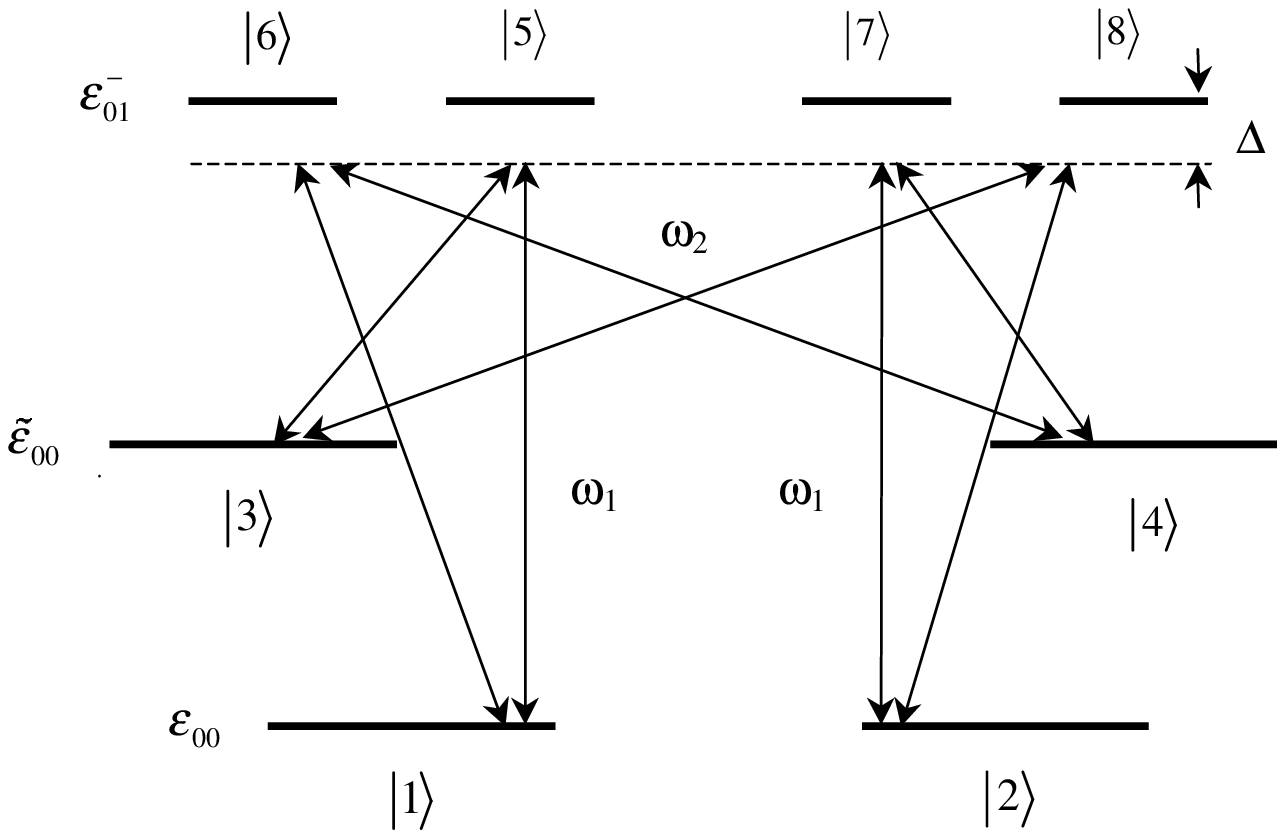}

\vskip 6mm

Fig. 3. Transition scheme connecting the states $\left| 1
\right\rangle $ and $\left| 2 \right\rangle $ through the use of
the auxiliary states $\left| 3 \right\rangle $, $\left| 4
\right\rangle $, $\left| 5 \right\rangle $, $\left| 6
\right\rangle $, $\left| 7 \right\rangle $, and $\left|
8\right\rangle $ in the two-photon resonant case, see Fig. 2.

\end{document}